# Writing Patterns Reveal a Hidden Division of Labor in Scientific Teams

**Authors:** Lulin Yang[1†], Jiaxin Pei[2†*], Lingfei Wu[1*]

**Affiliations:**

[1] School of Computing and Information, The University of Pittsburgh, 135 N Bellefield Ave, Pittsburgh, PA 15213

[2] Stanford Institute for Human-Centered Artificial Intelligence, 353 Jane Stanford Way, Stanford, CA 94305

†These authors contributed equally as co-first authors.

*Corresponding author. E-mail: liw105@pitt.edu (L. W.) and pedropei@stanford.edu (J. P.)

## Abstract

The recognition of individual contributions is central to the scientific reward system, yet coauthored papers often obscure who did what. Traditional proxies like author order assume a simplistic decline in contribution, while emerging practices such as self-reported roles are biased and limited in scope. We introduce a large-scale, behavior-based approach to identifying individual contributions in scientific papers. Using author-specific LaTeX macros as writing signatures, we analyze over 730,000 arXiv papers (1991–2023), covering over half a million scientists. Validated against self-reports, author order, disciplinary norms, and Overleaf records, our method reliably infers author-level writing activity. Section-level traces reveal a hidden division of labor: first authors focus on technical sections (e.g., Methods, Results), while last authors primarily contribute to conceptual sections (e.g., Introduction, Discussion). Our findings offer empirical evidence of labor specialization at scale and new tools to improve credit allocation in collaborative research.

## Main

Scientific breakthroughs today increasingly emerge from large teams, where the individual contributions of members are difficult to trace. Over the past decades, solo authorship has declined dramatically—from over 80% across all disciplines in the 1960s to much less by the 2010s: 50% in sociology, 26% in economics, and just 7% in computer science[1–3]. Landmark discoveries increasingly involve massive collaborations, such as the 2,800 researchers who mapped the human genome[4] and the 3,500 scientists who detected gravitational waves[5]. Indeed, coauthored papers have become the norm.

In response, authorship practices have evolved. Conventionally, author order has served as the primary indicator of contribution: the first author is assumed to contribute the most, followed by the second, and so on[6]. Yet this linear model fails to capture the diversity of contributions across team members. To address this gap, contribution statements—first adopted by prestigious journals like *Nature, Science,* and *PNAS*, and later by many others—aim to clarify who did what[7]. Frameworks like CRediT (Contributor Roles Taxonomy) have formalized this process across disciplines[8]. However, these systems remain self-reported, prone to bias, and limited to a small subset of journals. Therefore, despite their promise, a fundamental question remains: *How do scientific teams actually divide labor?*

This question is critical because the recognition of individual contributions is central to the scientific reward system. Effective credit allocation is essential for matching talent to effort and supporting innovation. In the 1960s, Robert Merton observed a persistent gap between contribution and recognition: citations—a proxy for recognition—disproportionately favored established scientists over junior peers, a pattern he termed the Matthew Effect[9]. He attributed this to selective memory in the scientific community: it is easier to remember collective achievements by crediting a few prominent individuals. This simplification often leaves the full scope of scientific labor unrecognized. Team science has intensified the issue: citations credit entire teams rather than individual members, and larger teams dilute credit by making it harder to determine who did what.

What patterns of labor division might we expect in science? Prior work suggests that innovation involves two distinct processes: ideation (e.g., developing hypotheses, framing questions) and execution (e.g., data analysis, experimentation)[10–13]. Inspired by this distinction, we hypothesize that scientific teams divide labor systematically across these principle components. Rather than assuming that contributions simply decline with author order, we ask whether specific types of labor cluster among particular team members.

To test this hypothesis, we introduce a large-scale, behavior-based approach to identifying writing contributions within scientific papers. We analyze LaTeX source code—the dominant format in STEM disciplines—to track author-specific macro usage. If an author previously used a macro that appears in any paper they authored—whether alone or with collaborators—we attribute the use of that macro to them[14]. By aggregating macro usage across sections, we estimate each author's share of writing contributions. Using 1.6 million LaTeX source files from arXiv.org, we identified contribution patterns for 583,817 scientists across 730,914 papers from 1991 to 2023.

We validate our estimates through four complementary strategies: comparing them to 1,276 self-reported author contributions in top journals (*Science*, *Nature*, *PNAS*, *PLOS ONE*); examining whether contribution declines with author order[6] (it does); confirming alignment with field-specific norms (e.g., alphabetical order in economics[15]); and correlating our estimates with detailed Overleaf editing histories. These validations support the robustness of our approach.

From this dataset, we identified 411,808 papers that included contributions to at least one major section, which were used to analyze the relationship between author order and section focus. A curated subset of 61,931 papers, in which the same author contributed to at least two major sections, was used to analyze the interdependence and clustering of sections. This curated subset enables meaningful comparisons across roles while preserving broad disciplinary and temporal coverage.

Our findings reveal a consistent and previously hidden division of labor in scientific writing. First authors focus on technical sections (e.g., Methods, Results), while last authors primarily contribute to conceptual sections (e.g., Introduction, Discussion). This role specialization is widespread across disciplines and team sizes, and aligns with self-reported contributorship data that link technical tasks with early-career researchers and conceptual tasks with senior scholars[16]. The phenomenon mirrors broader patterns of labor specialization across domains, from classical theories by Adam Smith[17] and Émile Durkheim[18] to contemporary studies of workplace inequality[19,20], where task differentiation can produce both efficiency and disparities in recognition. By uncovering structured task differentiation in scientific teams[10,11], our study provides new tools to analyze collaboration and inform more transparent, effective approaches to credit allocation in science.

**Results**

**Hidden Division of Labor in Scientific Teams**

We hypothesize that scientific teams exhibit a structured division of labor in writing, systematically allocating different types of work across members. Rather than assuming contributions simply decline with author order, we ask whether specific roles—such as conceptual versus technical work—cluster among team members. To examine this pattern, we estimate individual contributions by analyzing author-specific LaTeX macros across sections of scientific papers. Major sections (e.g., Introduction, Methods, Results) are identified using LaTeX \section commands. For each author, we normalize the number of unique macros in each section relative to their total macros in the paper, using this as a proxy for their section-level focus.

Figure 1 shows a clear trend: contributions shift from technical to conceptual sections as author order increases. Using LaTeX source files from 411,808 papers published between 1991 and 2023, we estimate regression models predicting section-specific contributions by author order. The results reveal that first authors contribute significantly more to technical sections such as Methods, Results, Experiments, and Preliminaries, whereas last authors disproportionately contribute to conceptual sections including Introduction, Discussion, Conception, and Acknowledgments.

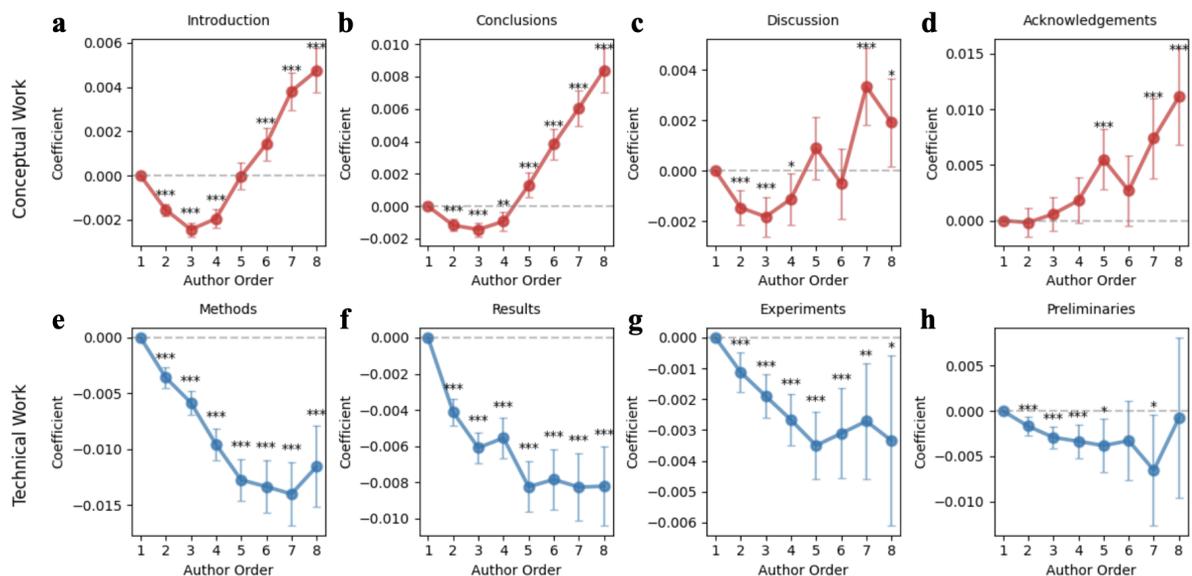

**Figure 1. Hidden Division of Labor in Scientific Teams.** We analyzed 411,808 papers published between 1991 and 2023, identifying sections based on LaTeX \section commands. To assess the relationship between author rank and their focal work—represented by the fraction of macros normalized within an author—we estimated regression models within each section. The coefficients represent the relative change in normalized author contributions compared to the first author, with 95% confidence intervals indicated by error bars. All models yielded $P < 0.01$, as denoted by stars in the figure. The trend shows a gradual shift from technical (**e-h**) to conceptual contributions (**a-d**) as author order increases.

**Conceptual and Technical Role Clustering**

In addition to author order patterns, we examined how contributions cluster across different sections within a paper. Using a curated subset of 61,931 papers in which the same author contributed to at least two major sections, we measured interdependence between sections based on conditional co-contribution probabilities. Specifically, we computed the conditional probability that an author contributing to one section also contributed to another, using LaTeX macro traces as behavioral evidence.

Figure 2 visualizes this section interdependence network. For consistency and to preserve the broadest coverage of papers, we standardized section labels to six major categories—selected from eight sections in the previous analysis—to ensure the highest data retention. Each node represents a major section, and directed edges reflect the conditional probability of shared authorship between them. Clustering analysis using Ward's hierarchical algorithm[21] reveals two distinct communities: conceptual sections (Introduction, Discussion, Conclusion) and technical sections (Preliminaries, Methods, Results).

This clear structural division indicates that authors tend to specialize within either conceptual or technical domains of a manuscript. These results extend our findings from author order and demonstrate that scientific writing reflects an underlying division of labor not just across authors, but also across the organizational structure of the paper itself.

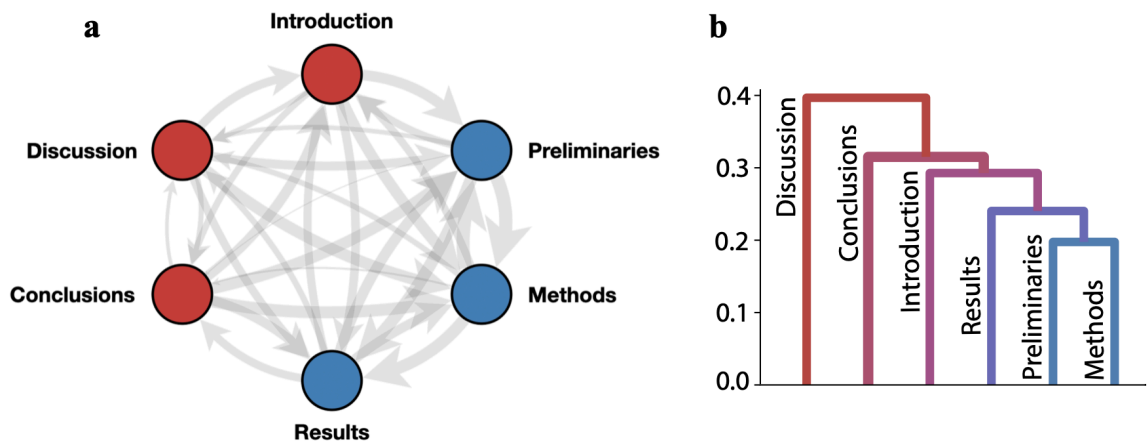

**Figure 2. Author Roles Revealed by Sectional Contributions.** We analyzed 61,931 LaTeX-authored papers published between 1991 and 2023 that include author contributions to at least two of the six major sections required for analysis. We identified section headings using LaTeX \section commands and standardized them to six major categories. We measured the interdependence between sections as the conditional probability of an author contributing to one section given that they had already contributed to another. (**a**) These probabilities are visualized as directed edges, with edge width proportional to the conditional probability. (**b**) Hierarchical cluster analysis[21] reveals two distinct communities: conceptual sections—Introduction, Discussion, and Conclusion (red nodes)—and technical sections—Preliminaries, Methods, and Results (blue nodes).

**Validation with Self-Reported Contributions**

To evaluate the validity of our macro-based estimates, we compared them to 1,276 self-reported author contribution statements from four journals—*Science*, *Nature*, *PNAS*, and *PLOS ONE*. We used OpenAlex metadata to identify arXiv preprints that were later published in these journals. For each identified paper, we retrieved the DOI and used a Python-based web crawler to extract contribution statements from publisher websites.

We benchmarked our estimates against these disclosures, assessing whether our method correctly identified writing contributors. The results show an average precision of 0.88 and a recall of 0.69, demonstrating that our macro-based method reliably infers writing roles at scale and aligns closely with self-reported contributions.

Table 1. Validation of author contributions inferred from LaTeX macros using self-reported data.

| Journal | Sample size N | Time period | Precision | Recall |
|---|---|---|---|---|
| PNAS | 599 | 2006-2023 | 0.90 | 0.73 |
| Nature | 265 | 2010-2023 | 0.78 | 0.65 |
| Science | 48 | 2018-2023 | 0.89 | 0.59 |
| Plos One | 362 | 2007-2023 | 0.92 | 0.65 |
| Total | 1,274 | 2006-2023 | 0.88 | 0.69 |

**Validation with Author Order Trends**

Author order is commonly used as a proxy for contribution, with the assumption that the first author has made the most substantial contribution[6]. To assess whether our method reflects this scientific authorship practice, we analyzed 411,808 papers in which at least one author contributed to a major section. We built regression models predicting writing contributions based on author order, controlling for team size.

As shown in Figure 3, contributions consistently decline with author order. First authors contribute significantly more than their coauthors. Notably, last authors contribute more than middle authors in large teams, aligning with their important conceptual role identified in the previous sections. These validations support the robustness of our estimates and confirm that our data aligns with established expectations of scientific authorship.

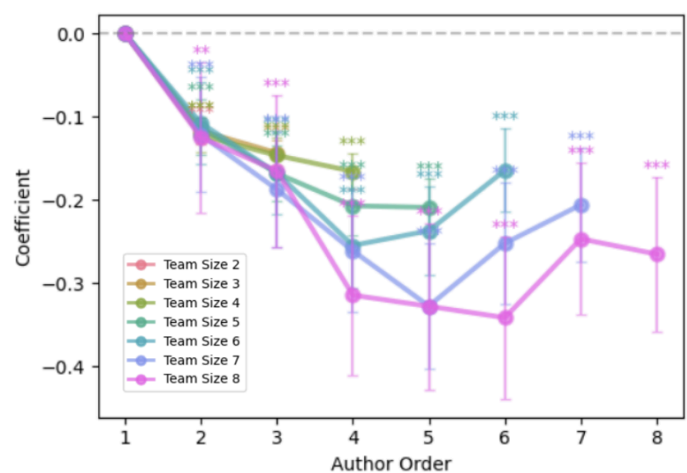

**Figure 3. Declining Contribution with Author Order.** We analyzed 411,808 papers published between 1991 and 2023, focusing on teams with up to eight members. To quantify the relationship between author order and contribution—measured by the number of unique LaTeX macros—we ran separate regression models for each team size. Coefficients represent the estimated change in contribution relative to the first author, with 95% confidence intervals shown as error bars. All models yielded $P < 0.01$, as indicated by stars.

**Disciplinary Consistency in Author Order**

Authorship conventions vary across disciplines, often reflecting different norms of contribution and credit. In computer science, for example, the first author typically leads the work, while the last author often acts as the senior advisor. In contrast, economics commonly uses alphabetical ordering, signaling presumed equality among coauthors[22].

To assess whether our contribution estimates align with these disciplinary norms, we ran separate regression models for computer science and economics papers, using author order to predict individual writing contributions. As shown in Figure 4, computer science papers display a strong decline in contributions with author order, while economics papers show a relatively flat distribution. These patterns affirm that our LaTeX-based approach captures discipline-specific authorship conventions, further validating our method.

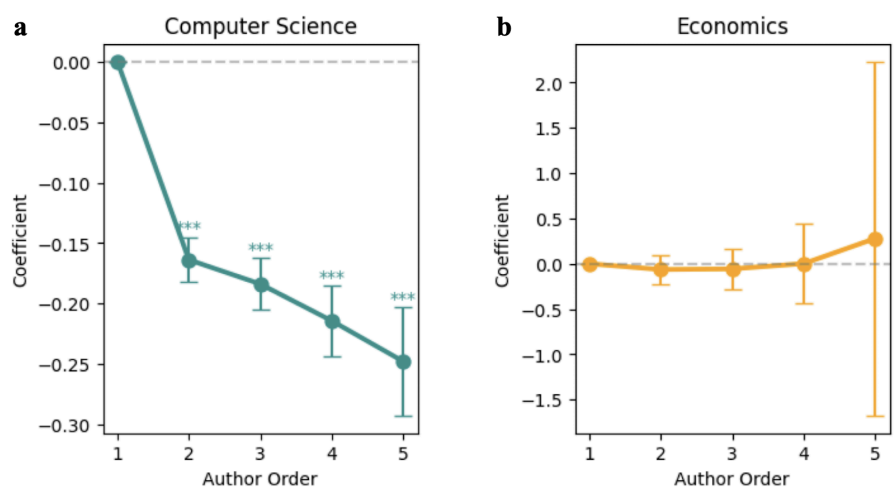

**Figure 4. Discipline-Specific Patterns in Author Order.** We analyzed 76,749 computer science papers and 1,194 economics papers published between 1991 and 2023, based on arXiv subject classifications. To examine the relationship between author order and contribution—measured by the number of unique LaTeX macros—we ran regression models within each discipline. Coefficients indicate the estimated change in contribution relative to the first author, with 95% confidence intervals shown as error bars. All models yielded $P < 0.01$, as indicated by stars.

**Validation with Overleaf Editing Histories**

To further validate our estimates, we analyzed detailed editing histories from Overleaf—a collaborative LaTeX writing platform widely used in STEM. We obtained editing logs for two multi-author papers, producing 14 author–paper records. For each paper, we manually counted each author's edits accumulated over the multi-month writing process and ranked them based on observed editing activity. This validation required intensive manual effort; notably, the most active contributor made 533 individual edits.

We then compared these rankings to our macro-based estimates. A Pearson correlation analysis revealed a moderate positive relationship (r = 0.50, p = 0.07), suggesting that our method captures meaningful signals of writing activity—even when compared to direct evidence from real-time collaborative environments. This behavioral validation complements our other benchmarks and reinforces confidence in our approach.

**Discussion**

This study provides the first large-scale behavioral analysis of writing contributions within scientific teams. By analyzing personalized LaTeX macros across more than 730,000 papers and validating our estimates against multiple benchmarks—including self-reported statements, author order trends, disciplinary norms, and Overleaf editing data—we reveal a structured division of labor in scientific writing. First authors focus on technical sections, while last authors tend to engage more with conceptual content. Contributions also cluster into technical and conceptual roles at the section level, revealing hidden specialization not only among team members but also across the internal structure of scientific papers.

These patterns challenge the simplistic assumptions underlying traditional authorship conventions. While first authors contribute the most, last authors—especially in large teams—contribute significantly to conceptual framing, which may explain why they receive disproportionate recognition under systems influenced by the Matthew Effect. This role asymmetry creates challenges for equitable credit allocation, particularly for junior researchers performing technical work. If early-career scientists are perceived primarily as "muscle" rather than the "brain" of a project[10,12], this could limit their trajectory toward intellectual independence and senior authorship opportunities[23]. As team sizes grow and temporary academic positions proliferate[6], these dynamics may exacerbate career disparities[24].

Our findings have direct implications for research evaluation and science policy. Current models of credit allocation often rely on author order or self-reported statements, both of which mask the true diversity of contributions. By surfacing behavioral traces of labor division, our approach supports more transparent and evidence-based practices for evaluating scientific work. It may also help scholars make more informed decisions about their team roles, authorship strategies, and collaborations.

Future work can extend this framework to natural language patterns or other collaborative traces, allowing for broader coverage beyond LaTeX-intensive fields. Researchers may also use this method to assess whether certain writing roles correlate with scientific impact, career outcomes, or funding success. Ultimately, recognizing how labor is divided in science is essential for designing institutions that reward both creativity and execution—and for ensuring that credit flows fairly in the age of team-based research.

**Methods**

**Data Collection**

We collected 1,600,627 LaTeX source files from arXiv.org, spanning papers submitted between 1991 and 2023 by 2,012,092 unique authors. These files cover major STEM fields where LaTeX is the dominant authoring format. Each file was parsed to extract section-level structure and user-defined macros. We filtered the dataset to include 730,914 multi-authored papers from which at least one author's contribution could be inferred.

**Author Macro Attribution**

We used author-specific LaTeX macro histories to infer writing contributions. For each author, we constructed a database of unique macros—user-defined commands that serve as writing signatures. If a macro from an author's prior work appeared in a paper they coauthored, we attributed that instance to the author. When macros overlapped across authors, credit was shared. Each author's contribution share was computed by normalizing their unique attributed macros over the total number of attributed macros in the paper. Figure 5 provides a concrete illustration of this process using the writing histories of two authors.

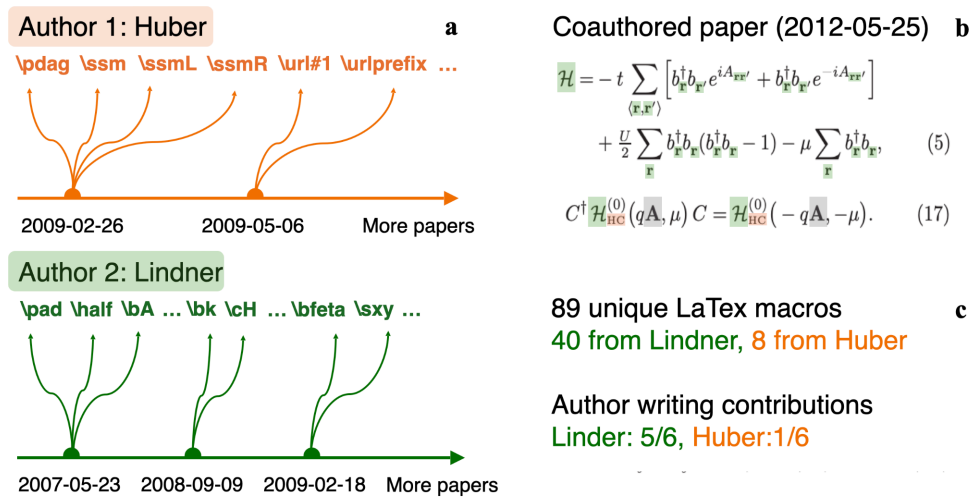

**Figure 5. Estimating Author Contributions via Macro Usage and Collaborative Traces.** (**a**) Macro usage histories for two authors, Huber and Lindner, based on their publications. Huber authored 42 papers with 26 unique macros; Lindner authored 48 papers with 195 unique macros. (**b**) In a 2012 coauthored paper, both authors' macros appear in the LaTeX source. Using their personal macro databases, we attribute 40 macros to Lindner and 8 to Huber. (**c**) These counts are normalized to estimate writing contributions of 5/6 for Lindner and 1/6 for Huber.

**Section-Based Analysis**

We extracted major section titles using the LaTeX \section commands and mapped them to canonical categories (e.g., Introduction, Methods, Results, Discussion). We identified 411,808 papers that included author contributions to at least one major section, which were used to analyze the relationship between author order and section focus. For co-contribution and clustering analysis, we used a curated subset of 61,931 papers in which the same author contributed to at least two major sections. A simplified six-section taxonomy was applied to preserve maximal coverage in network analysis.

**Validation Benchmarks**

We validated our method using four complementary strategies, each aligning our macro-based estimates with established expectations. We compared our results to 1,276 self-reported author contribution statements from papers published in *Science*, *Nature*, *PNAS*, and *PLOS ONE*, extracted via a Python web crawler. We tested whether contributions declined with author order across 411,808 papers, and whether patterns in computer science (n = 76,749) and economics (n = 1,194) aligned with field-specific authorship conventions. Finally, we compared our estimates with Overleaf edit histories for two multi-author papers, observing a moderate correlation (r = 0.50, *p* = 0.07).

**Clustering Analysis**

To study section-level labor specialization, we computed conditional co-contribution probabilities—i.e., the likelihood that an author contributing to one section also contributed to another. These probabilities were modeled as a directed network and analyzed using Ward's hierarchical clustering algorithm[21] to detect conceptual vs. technical role clusters.

**Limitations and the Collaborative Learning Assumption**

Our method relies on a key assumption we term the "collaborative learning" assumption: authors tend to adopt macros from their collaborators and continue to use them. We find evidence supporting this assumption. For example, in a 2012 coauthored paper, Lindner adopted Huber's macro "\newcommand{\ssm}{\scriptscriptstyle\rm}"—previously used in Huber's solo work—and reused it in subsequent papers. This reuse illustrates the collaborative learning dynamic in practice.

While generally supported, this assumption may not always hold, potentially introducing bias against junior authors with limited macro histories. Moreover, as our method depends on the LaTeX source code, it cannot be directly applied to fields where LaTeX is not the standard authoring tool. However, the underlying principle—inferring contribution from behavioral writing traces—may inspire broader applications.

**Data availability**

The dataset used in this study has been deposited in Zenodo and is available at https://zenodo.org/records/15243518.

**Code availability**

All notebooks necessary to reproduce the results are available on GitHub at https://github.com/AnomieYang/Writing-Patterns-Observations.

**Acknowledgments.** We are grateful for support from the National Science Foundation grant SOS: DCI 2239418 (L.W.). This work was conducted while J.P. was supported by startup funding to L.W. from the University of Pittsburgh.

**Author contributions.** L.W. led the conceptualization and manuscript writing. J.P. developed the initial dataset. J.P. and L.Y. conducted the data analysis, implemented the models, and contributed to manuscript editing. All authors contributed to discussions.